\documentclass[fleqn,usenatbib]{mnras}

\usepackage{newtxtext,newtxmath}
\usepackage{wasysym}
\usepackage{orcidlink}
\usepackage[T1]{fontenc}
\usepackage{graphicx}
\usepackage{xcolor}
\newcommand{\orcid}[1]{\href{https://orcid.org/#1}{\textcolor[HTML]{A6CE39}{\aiOrcid}}}
\DeclareUnicodeCharacter{2212}{-}
\DeclareRobustCommand{\VAN}[3]{#2}
\let\VANthebibliography\thebibliography
\def\thebibliography{\DeclareRobustCommand{\VAN}[3]{##3}\VANthebibliography}

\usepackage{graphicx}	% Including figure files
\usepackage{amsmath}	% Advanced maths commands

\title[Narrow absorption line Outflow in Seyfert 1 galaxy  J1429+4518]{Narrow absorption line Outflow in Seyfert 1 galaxy  J1429+4518:\\ Outflow's distance from the central source and its energetics}

\author[M. Dehghanian et al.]{
M. Dehghanian,$^{1}$\orcidlink{0000-0002-0964-7500} \thanks{E-mail: dehghanian@vt.edu}
N. Arav,$^{1}$\orcidlink{0000-0003-2991-4618}
D. Byun$^{1}$\orcidlink{0000-0002-3687-6552}
G. Walker$^{1}$\orcidlink{0000-0001-6421-2449}
and M. Sharma$^{1}$\orcidlink{0009-0001-5990-5790}
\\
$^{1}$Department of Physics, Virginia Tech, Blacksburg, VA 24061, USA\\
}

\date{Accepted 2023 November 24. Received 2023 October 16; in original form 2023 August 25}

\pubyear{2023}

\begin{document}
\label{firstpage}
\pagerange{\pageref{firstpage}--\pageref{lastpage}}
\maketitle

\begin{abstract}
In the HST/COS spectrum of the Seyfert 1 galaxy 2MASX 
J14292507+4518318, we have identified a narrow 
absorption line (NAL) outflow system with a velocity 
of $-$151 km s$^{−1}$. This outflow exhibits 
absorption troughs from the resonance states
of ions like \ion{C}{iv}, \ion{N}{v}, \ion{S}{iv}, and \ion{Si}{ii}, as well
as excited states from \ion{C}{ii}$^{*}$, and 
\ion{Si}{ii}$^{*}$. Our investigation of the outflow involved measuring ionic
column densities and conducting photoionization analysis.  These yield the total column density of the outflow to be estimated as $\log N_{H}$=19.84 [cm$^{-2}]$, its ionization parameter to be $\log U_{H}$=-2.0 and its electron number density equal to $\log n_{e}$= 2.75[cm$^{-3}$]. These measurements enabled us to determine the 
mass-loss rate and the kinetic luminosity of the outflow system to be $\Dot{M}$=0.22~[$M_{\astrosun}~yr^{-1}$] and $\log \Dot{E_{K}}$=39.3~[erg s$^{-1}$], respectively.  We have also measured the location of the outflow system to be at  $\sim$275 pc from the central source. This outflow does not contribute to the AGN feedback processes 
 due to the low ratio of the outflow's kinetic luminosity to the AGN's Eddington luminosity ($\Dot{E_{K}}/{L_{Edd}}\approx 0.00025 \%$). This outflow is remarkably similar to the two bipolar lobe outflows observed in
the Milky Way by XMM-Newton and Chandra.
\end{abstract}


%%%%%%%%%%%%%%%%%%%%%%%%%%%%%%%%%%%%%%%%%%%%%%%%%%

%%%%%%%%%%%%%%%%% BODY OF PAPER %%%%%%%%%%%%%%%%%%

\section{Introduction}
Absorption outflows from active galactic nuclei (AGN) are frequently suggested as potential contributors
to the AGN feedback processes.
Various
studies have discussed this 
idea, e.g. \citet{silk98, scan04,yuan18,vayner21, he22}. Absorption lines found in the rest-frame UV spectra are typically grouped into three main categories based on their characteristics: broad absorption lines (BALs) which have a width of $\geq$2000 km~s$^{-1}$, narrow absorption lines (NALs) with a width of $\leq$500 km~s$^{-1}$, and a middle subgroup known as mini-BALs \citep{itoh20}.
When it comes to BALs, it is clear that the absorbing gas is connected to the AGN \citep{vest03} and could indeed have a noticeable impact on suppressing the star formation rate within the host galaxies \citep{chen22}; however, the nature of NAL outflows is less understood. This is due to the challenge of discerning intrinsic NALs (associated NALs) from those NALs that are unrelated to the quasars (intervening NALs). Intervening NALs can originate from various sources, including intervening galaxies, intergalactic clouds, Milky Way gas, or gas within the host galaxies of the quasars \citep{misa07a}. Several studies (e.g. \citet{misa07a,chen13,chen18a,chen20}) have highlighted the point that variations in the absorption lines serve as a robust indicator to distinguish between narrow absorption lines arising from the outflow system, or those called intervening NALs. These variations in absorption lines typically arise from changes in the column density of absorbing gas as part of the intrinsic mechanism. 

\cite{shen12} and \cite{chen18b} explain that a velocity of 3000 km~s$^{-1}$ is often used to limit associated NALs; however, it should also be noted that some NAL outflows with velocity being obviously larger than 3000 km~s$^{-1}$ have been observed \citep{ham11,chen13}. Using large samples of the SDSS quasar \ion{C}{iv} and \ion{Mg}{ii} narrow-line absorptions, \cite{chen17} discuss that the velocity boundary is associated with the quasar luminosity. In the study referred to, it is proposed that velocity thresholds of 4000 km~s$^{-1}$ be used to identify quasar CIV-associated NALs and velocity thresholds of 2000 km~s$^{-1}$ be employed to identify quasar MgII-associated NALs.

Despite the fact that NALs have received less attention than BALs, they hold great potential as a useful tool for investigating the physical properties of outflows because of these two significant reasons \citep{misa07}:
\begin{itemize}
    \item NALs do not have the problem of self-blending, which is defined as blending blue and red components of doublets like \ion{C}{iv} $\lambda\lambda$ 1548,1551\AA.
    \item NALs appear in a larger variety of AGNs, whereas BALs are mostly detected in radio-quiet quasars. 
\end{itemize}

This paper identifies and analyzes the NAL outflow system of the AGN 2MASX J14292507+4518318 (hereafter J1429+4518) using the observations performed by the Hubble Space Telescope (HST) in 2021. 
For this object, the intrinsic nature of the NAL is deduced from the presence of several excited state absorption narrow lines observed in the spectrum. 
The paper's structure is as follows:

\noindent In the next Section, we described the observations and data acquisition of the target. It is followed by Section~\ref{sec:anal}, which includes the analysis, and explains the methods used in the paper. Section~\ref{sec:anal} also details the approaches used to calculate the ionic column densities of the NALs. Calculations of the electron number density and location of the outflow, along with the best photoionization solution, are all included in Section~\ref{sec:anal}. 
Part of Section~\ref{sec:anal} is dedicated to finding the central source's mass and the AGN's Eddington luminosity. These calculations are followed by estimating the energetics of the outflow system. 
Section~\ref{sec:disc} discusses the stability of the outflow and indicates the importance of the NALs in general. In this Section, we have also compared our NAL outflow system with two lobe outflows detected in the Milky Way. Finally, Section~\ref{sec: con} gives a summary of what we have done here and concludes the paper.

Here we adopted a cosmology with h= 0.696, $\Omega$= 0.286, and
$\Omega_\nu$ = 0.714 \citep{benn14}. We used the Python astronomy
package Astropy \citep{astro13,astro18} for our
cosmological calculations, as well as Scipy \citep{virt20},
Numpy \citep{harr20}, and Pandas \citep{reba21} for most of our numerical computations. For our plotting purposes, we used Matplotlib \citep{hunt07}.
%%%%%%%%%%%%%%%%%%%%%%%%%%%%%%%%%%%%%%%%%%%%%%
\section{Observations} 
\begin{figure*}
\centering
\includegraphics[width=7 in]{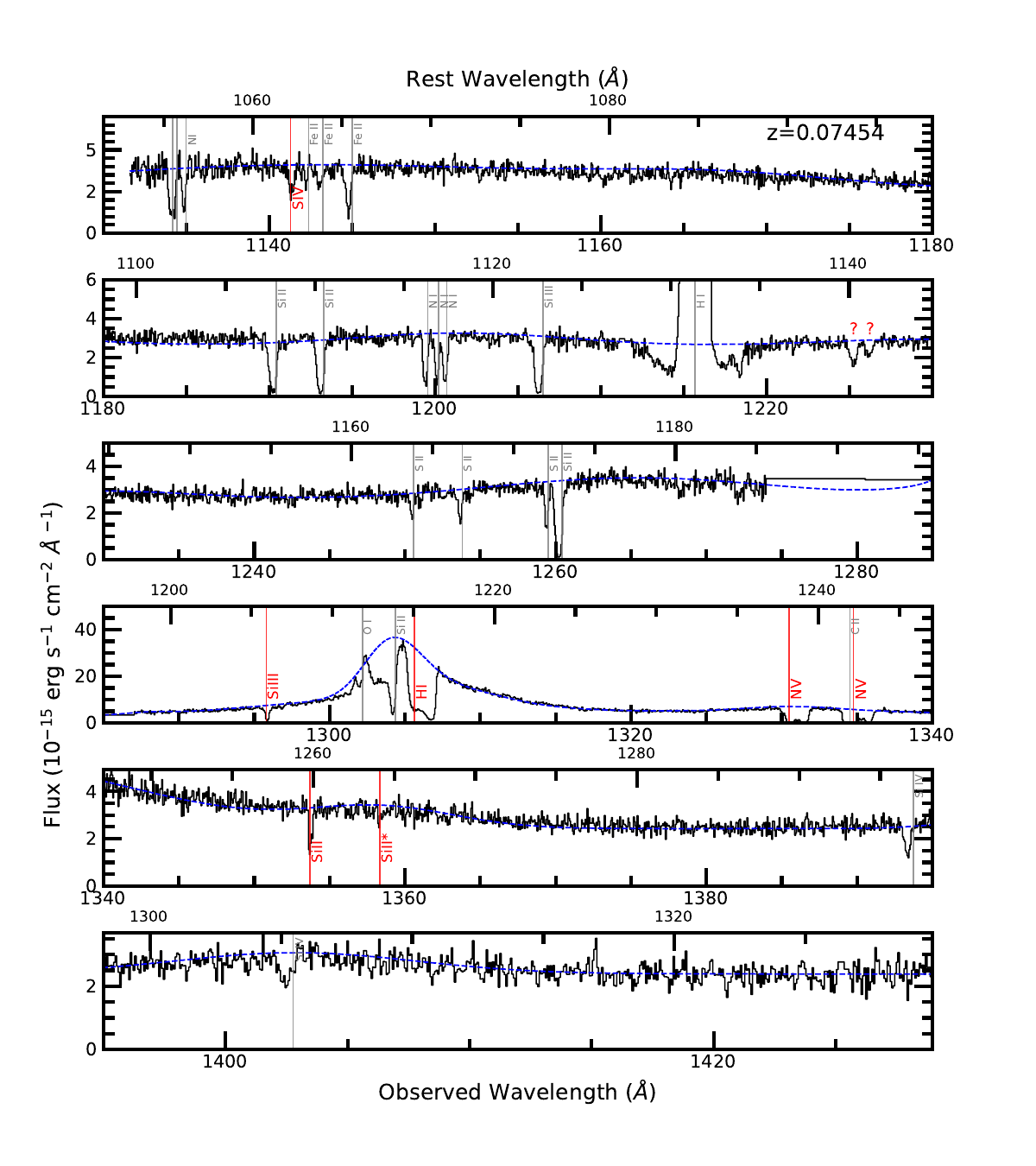}
      \caption{The spectrum of J1429+4518 observed by the HST/COS in Sep 2021\citep{prop}. The absorption features of the outflow system with a velocity of --151 km/s are marked with red lines, while the grey lines show the absorption from the interstellar medium (ISM). The dashed blue line shows our continuum emission model. The question marks indicate some absorption lines that we could not identify (we confirm they are not CIV resulting from an intervening system).} 
         \label{fig1}
\end{figure*}

J1429+4518 is a Seyfert 1 galaxy, located at Redshift
z=0.07454, with J2000 coordinates at 
RA=14:29:25.07 and DEC=+45:18:31.93. The
galaxy was observed using the HST 
Cosmic Origins Spectrograph (HST/COS) on three separate
occasions: September 19th 2021, December 20th of 2021, 
and December 21st of the same year, as part of HST Proposal 
16301 \citep{prop}. The first observation utilized the G130M 
grating, while the subsequent two observations used
G160. For this study, we excluded the 
observation taken on December 21st of 2021, as it covered
the same wavelength range as the other December 
observation but had a shorter exposure time. The 
spectral data for both observations are displayed in 
Figs~\ref{fig1} and \ref{fig2} .

After obtaining the relevant data from the Mikulski 
Archive for Space Telescopes (MAST), we detected an 
outflow at a velocity of v=~--151 km s$^{-1}$, with blue-shifted ionic 
absorption lines denoted by red vertical lines in Figures~\ref{fig1} $\&$ \ref{fig2}. Among 
the identified absorption lines, several well-known 
resonance doublets, such as \ion{C}{iv} and \ion{N}{v}, were observed.
Additionally, we identified absorption lines 
arising from excited states, such as \ion{C}{ii}$^{*}$ and \ion{Si}{ii}$^{*}$.

\subsection{Redshift considerations}
\label{subsec:red}
The redshift assigned to the object (z=0.07454) is the sole value documented in the SIMBAD Astronomical Database \footnote{SIMBAD Astronomical Database: \url{http://simbad.cds.unistra.fr/simbad/}} and is derived from optical data analysis. Despite the NASA/IPAC Extragalactic Database (NED)\footnote{NED:\url{https://ned.ipac.caltech.edu/}} reporting seven different redshift values (within the range of z=0.0745 to z=0.0748) for this object, their preferred redshift in their databases is also documented as 0.07454. Furthermore, the SDSS Science Archive Server (SAS)\footnote{SAS:\url{https://dr18.sdss.org/}} confirms this redshift for the same object. In the interest of maintaining consistency with these databases, we have adopted the same redshift value. It is essential to acknowledge the inherent uncertainties associated with this value:

\noindent To ensure that we have picked a reliable redshift and to mitigate the potential larger uncertainties associated with \ion{C}{iv} emission line models \citep{chen19,shen11}, we opted to utilize the SDSS data directly and perform modeling on [\ion{O}{iii}]$\lambda$5007\AA. For this purpose, we have used the optical spectrum obtained in March 2003 (MJD 52728) and used two Guassians to fit [\ion{O}{iii}] emission line. Our modeling yielded to a redshift of 0.0747, a value
only 0.00016 larger than the NED estimate. However, this will
increase the velocity of the outflow to be v= –195 km~s$^{-1}$. Taking
all these considerations into account, we utilize the initially assumed
redshift throughout this paper, and we assume an uncertainty of $\Delta$v$\sim$44
km~s$^{-1}$ when needed.

\section{Analysis}
\label{sec:anal}
\subsection{Ionic Column Densities}
\label{ionCD}
Determining the ionic column densities
(N$_{ion}$) of the outflow absorption lines is crucial 
for understanding the physical properties of the outflow system. While there are two methods to estimate the ionic column densities, we employed "the apparent optical depth" (AOD) method for all detected absorption lines since all of the identified doublets appear to be saturated (see Sections~\ref{subsec:c4}). In this method, the main assumption is that the outflow 
uniformly and fully covers the 
source \citep{spit68,sava91}. When using the AOD method, one can infer the column density via the equations~\ref{eq-1}$\&$ \ref{eq0} below  \citep{spit68,sava91}:

\begin{equation}
I(\lambda)=I_{0}(\lambda) e^{-\tau(\lambda)}, \label{eq-1}
\end{equation}

\noindent in which $I(\lambda)$ and $I_{0}(\lambda)$ are the intensities with and without the absorption, respectively. $\tau(\lambda)$ is the optical depth of the absorption trough. Then, the ionic column densities can be measured using equation~\ref{eq0} \citep{arav01}:
%%%%%%%%%%%%%%%%%%%%%%%%%
\begin{equation}
N_{ion}=\frac{3.7679\times 10^{14}}{\lambda_{0}f_{ik}}\times \int \tau(\nu)~d\nu 
 [cm^{-2}],
\label{eq0}
\end{equation}
%%%%%%%%%%%%%%%%%%%%%%%
\noindent where $\lambda_{0}$ is the transition's wavelength and $f_{ik}$ is the oscillator strength of the transition. 
For more details regarding the AOD method, please see \cite{arav01}, \cite{gabel03}, and \cite{byun22c}. The other method is called the partial covering (PC) method, and it will not be used in this paper. For a detailed explanation of both methods and to better understand the logic behind them and get a sense of their 
mathematics, please refer to \cite{barl97,arav99a,arav99b, kool02, arav05,borg12a,byun22b, byun22c}.

%Both methods are explained below:
%\begin{itemize}
%\item The partial covering (PC) method: This method is used when we detect two lines arising from a same-energy lower state such as \ion{C}{iv} or \ion{Si}{iv}. When observing these lines in absorption, we expect the blue trough to be deeper than the red one (any ratio between 1:1 and 2:1) if the optical
%depth is relatively low (this will be the "linear" part of the curve of growth). Indeed this method assumes that the covering factor is less than unity \citep{barl97,arav99a,arav99b}. The ionic column densities resulting from the PC method are rather considered accurate measurements, not upper/lower limits \citep[e.g.][]{kool02, arav05,borg12a,byun22b}.
%\item The apparent optical depth (AOD) method: In this method, the main assumption is that the outflow 
% uniformly and fully covers the 
 %source \citep{sava91}. The AOD enables us to calculate an upper or lower limit for the ionic column density but not an exact measurement. Following the discussion in \cite{arav01} and \cite{gabel03}, we employ this method whenever we have a singlet absorption line. We also use AOD method for the doublets such as \ion{C}{iv} and \ion{Si}{iv} whenever the lines reach saturation (optical depth line ratios $\sim$1:1). Notably, a saturated line doesn't necessarily appear black; it implies that the outflow system only partially covers the background source, allowing some photons to pass through. 
%\end{itemize}
 Below we 
explain how various absorption lines were treated
and why we applied the AOD method approach in each case.
For each absorption line that we were interested in, we 
used the redshift of the 
outflow (z$_{\textrm{outflow}}$) to transfer the spectrum from 
wavelength space to velocity space (please refer 
to Fig.~\ref{fig3}). As shown by the dashed lines in 
Fig.~\ref{fig3}, we have chosen a velocity range of -210 km~s$^{-1}$ to -90 
km~s$^{-1}$ as our integration range when performing column density 
calculations. This region was chosen based on the centroid velocity and the absorption trough width of \ion{Si}{iv}. Since all the absorption lines result from the same outflow system, we use the same integration range for all.
%%%%%%%%%%%%%%%%%%%%%%%%%%%%%%%%%%
\subsubsection{\ion{C}{iv}, \ion{Si}{iv}, and \ion{N}{v} doublets}
\label{subsec:c4}
\label{subsec:n5}
The \ion{C}{iv} $\lambda \lambda$1548, 1551\AA,\   \ion{Si}{iv} $\lambda \lambda$1394, 1403\AA \ and \ion{N}{v} $\lambda\lambda$1239, 1243\AA \ absorption lines are prominent spectral features observed in AGN outflows. All of these lines are easily detectable in the spectrum of J1429+4518 (see Figs~\ref{fig1} and ~\ref{fig2}) indicating the presence of an absorption outflow system. However, while these lines are recognized as doublets (and expected to be treated using the PC method), they are all very saturated as their blue-to-red trough ratio is very close to 1:1 for all three cases. For this reason, in our calculations, we adopted the AOD results of the red doublet (since it has the lower oscillator strength) and considered them a lower limit to the total ionic column density for each species.
Note that for the case of \ion{N}{v}, the high-velocity wing of the red doublet is actually Galactic \ion{C}{ii} $\lambda$1334\AA\ absorption.
%%%%%%%%%%%%%%%%%%
\subsubsection{Ly$\alpha$}
As shown in Fig.~\ref{fig3}, Ly$\alpha$ is also suspected of being highly saturated since it has a very large optical depth. For this reason, while we use the AOD method to calculate its ionic column density, we take it as a lower limit and not an accurate measurement. 
%%%%%%%%%%%%%%%%%%%%
%%%%%%%%%%%%%%%%%
\subsubsection{\ion{C}{ii} $\lambda$ 1334 \AA \ and \ion{C}{ii}$^{*}$ $\lambda$ 1336
 \AA}
 \label{subsec:c2}
Since these two lines are from different energy levels, we employ the AOD method to calculate the column density for each line individually. However, as depicted in Fig.~\ref{fig3}, the flux ratio of \ion{C}{ii} to \ion{C}{ii}$^*$ appears to be approximately 1:1. While it is 
evident that the lines are not fully saturated (optical depth is not very large), a 1:1 ratio might have resulted from a mild saturation effect. For this reason, we proceed to utilize their column density resulted from the AOD method as a lower limit, not an accurate measurement.
In later steps, we use the ratio of $\frac{\textrm{N}_{ion}(\ion{C}{ii}^{*})}{\textrm{N}_{ion}(\ion{C}{ii})}$ to extract a lower limit for the electron number density of the outflow.
%%%%%%%%%%%%%%%%%%%%%%%%
\subsubsection{\ion{Si}{iii} $\lambda$ 1206\AA}
As marked in Fig.~\ref{fig1}, \ion{Si}{iii} $\lambda$ 1206\AA\  is easily identifiable. Since this line is a singlet, we use our usual AOD method to infer its ionic column density. However, since it appears to be very deep (see Fig.~\ref{fig3}) and hence probably saturated, we consider the results as a lower limit, not an actual measurement.
%%%%%%%%%%%%%%%%%%%%%%%%

% \subsubsection{\ion{Fe}{ii} $^{*}$ $\lambda$ 1636\AA}
% While, as shown in Fig.~\ref{fig3}, it seems there are some absorption features at $\sim$1636$\AA$, we do not calculate a column density for the excited state of \ion{Fe}{ii} absorption line. Given the noise of the data and the negligible oscillator strength of this line, we ignore this absorption-like depth in the spectra. The same discussion holds for the \ion{Si}{ii}$^{*}$ $\lambda$ 1533 $\AA$. Furthermore, please notice that this later line has an oscillator strength much smaller than \ion{Si}{ii}~1526\AA meaning that we do not expect to see \ion{Si}{ii} $^{*}$ $\lambda$~1533\AA when \ion{Si}{ii}1526\AA is not detected.

%%%%%%%%%%%%%%%%%%%%%%%
\subsubsection{\ion{Si}{ii} $\lambda$1260 \AA\ and \ion{Si}{ii}$^{*}$ $\lambda$1265
 \AA}
 \label{subsec:si2}
Fig.~\ref{fig3} shows that both resonance and excited state absorption lines of \ion{Si}{ii} ion are detectable.  Since the \ion{Si}{ii} trough is significantly shallower than the \ion{Si}{iii} trough, we cautiously accept its AOD measurement as a real measurement as opposed to a lower limit. The fact that  \ion{Si}{ii}$^{*}$ is shallower than \ion{Si}{ii} suggests that we can use their ionic column
densities ratio to estimate the electron number
density of the outflow  \citep{arav18}. Later in Section~\ref{subsec:eden} we discuss this process in more detail.
%%%%%%%%%%%%%%%%%%
%\pagebreak

%%%%%%%%%%%%%%%%%%%%%%%%%%%%%%%%%%%%%
\begin{figure*}
\centering
\includegraphics[width=7 in]{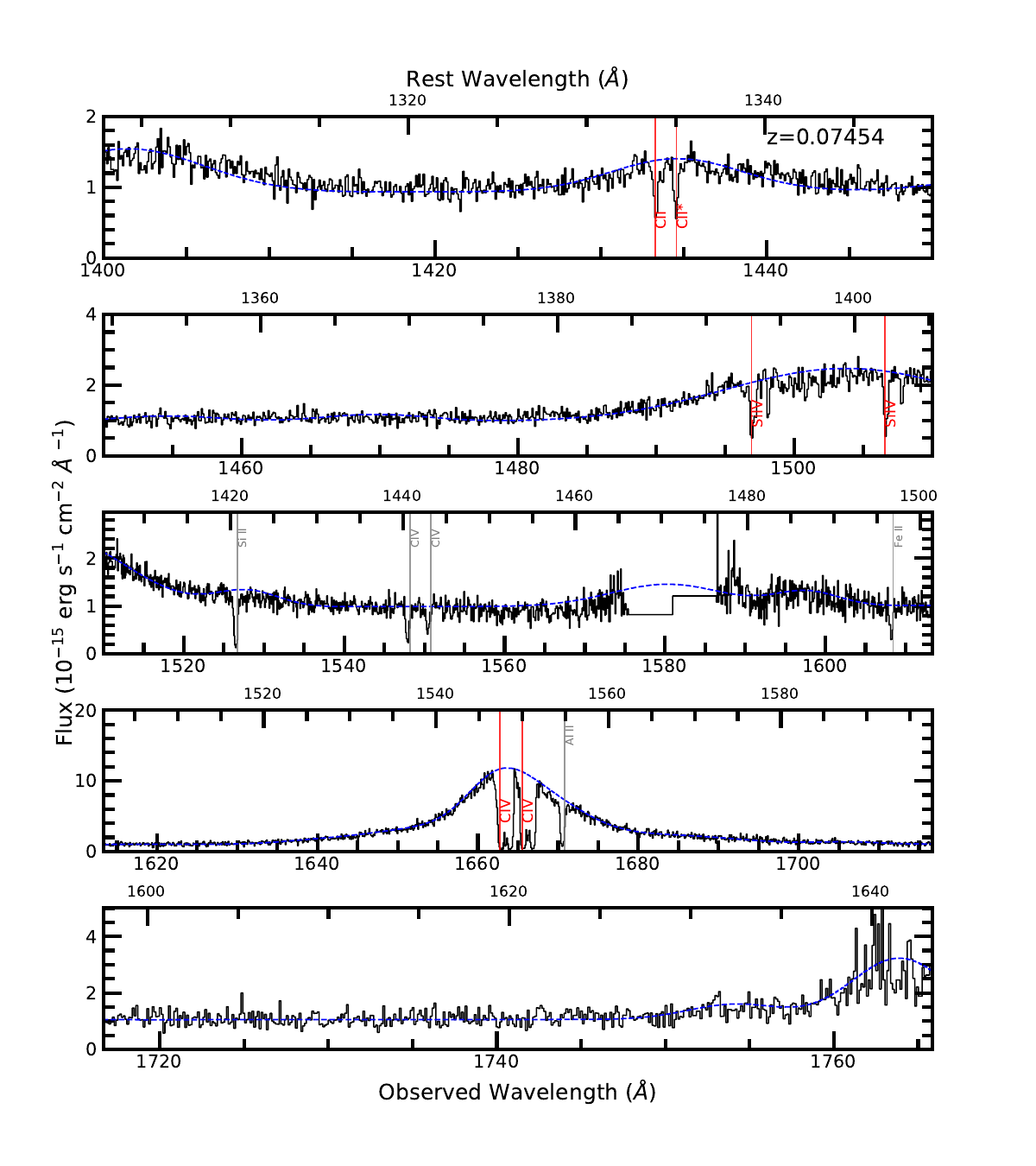}
      \caption{Same as Fig.\ref{fig1} but the data are from HST/COS observations in Dec 2021 \citep{prop}. Note that since the \ion{C}{iv} lines are black at the bottom, the covering fraction of the NAL outflow is estimated to be 1.0.}
         \label{fig2}
\end{figure*}
%\pagebreak
%%%%%%%%%%%%%%%%%%%%%%%
\begin{figure*}
\centering
\includegraphics[width=7 in]{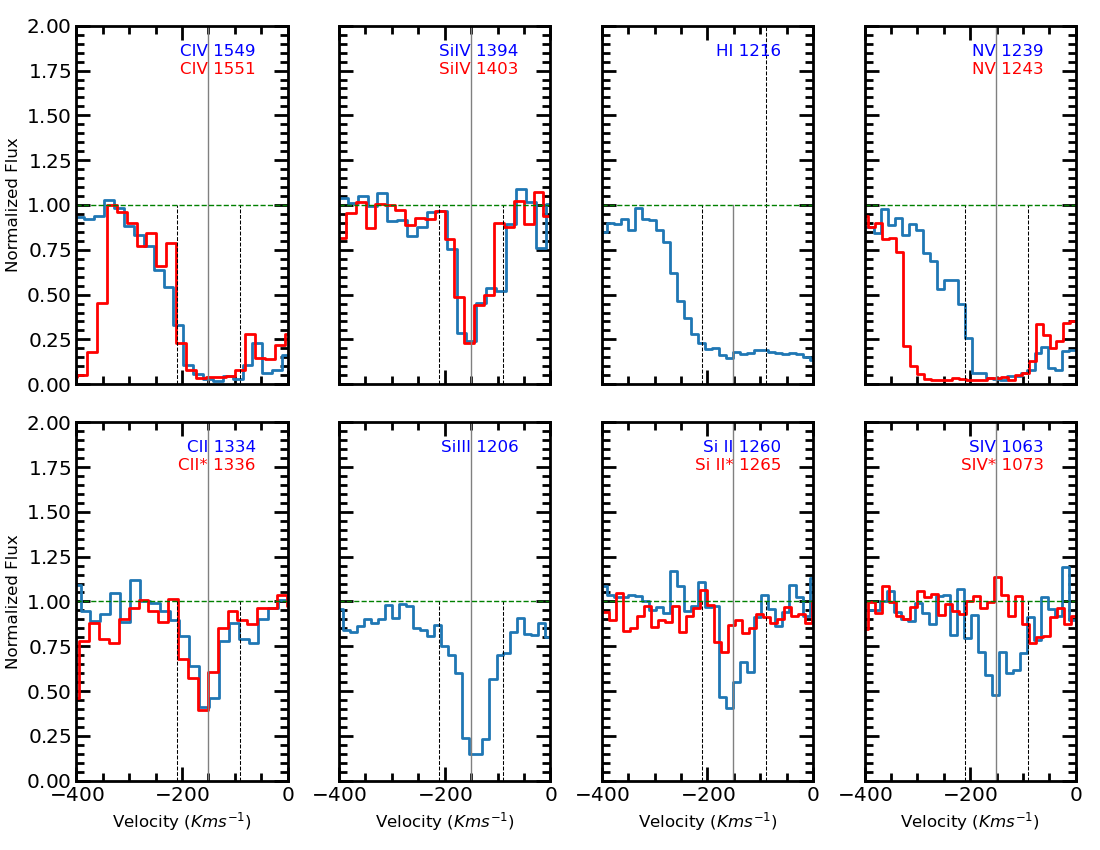}
      \caption{Normalized flux versus velocity for blue-shifted absorption lines detected in the spectrum of J1429+4518. The horizontal green dashed horizontal line shows the continuum level,
and the vertical black dashed lines show the region between −210 and −90
km s$^{−1}$. The vertical solid black line indicates the centroid velocity of v$_{\textrm{centroid}}$=--151 kms$^{-1}$.}
         \label{fig3}
\end{figure*}
%%%%%%%%%%%%%%%%%%%%%%%%%%
%%%%%%%%%%%%%%%%%%%%%%%%%%%%%%%%%%%
\subsubsection{\ion{S}{iv} $\lambda$1063 \AA \ and \ion{S}{iv}$^{*}$ $\lambda$1073
 \AA}
 \label{subsec: s4}
While we were able to accurately measure the column density of the \ion{S}{iv} using the AOD method, \ion{S}{iv}$^{*}$ is absent (Fig.~\ref{fig3}). For this reason, we used a Gaussian fit based on the \ion{S}{iv} absorption line and estimated an upper limit for \ion{S}{iv}$^{*}$ ionic column density. Then the ratio of N$_{ion}$(\ion{S}{iv}$^{*}$) to N$_{ion}$(\ion{S}{iv}) can be used to extract an upper limit for the electron number density of the outflow system.

\begin{figure}
\includegraphics[width=\columnwidth]{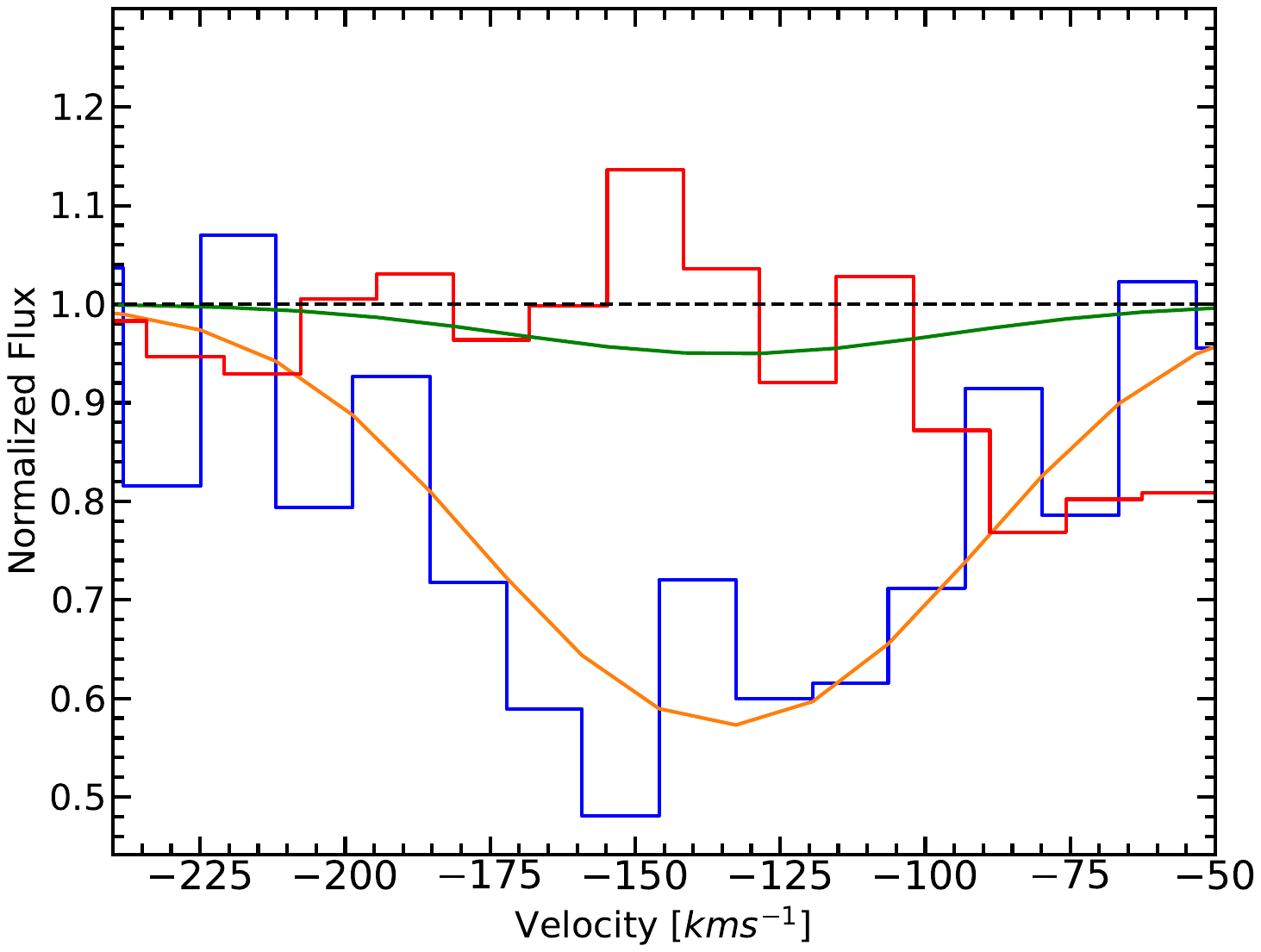}
      \caption{Gaussian modeling (green) of the \ion{S}{iv}$^{*}$$\lambda$1073
 \AA\  absorption trough, created by fitting 
\ion{S}{iv} $\lambda$1063 \AA\  line (orange) and using its shape as a template, where we have allowed the optical depth to adjust within 3$\sigma$ error range. The horizontal dashed
line shows the continuum level.}
         \label{fig4}
\end{figure}

Fig.~\ref{fig4} illustrates the Gaussian fit for both the resonance and excited states of the \ion{S}{iv} line. It is important to note that we modeled the \ion{S}{iv}$^{*}$ state using the best-fit parameters obtained for the \ion{S}{iv} resonance line as the template. The depth of the curve for \ion{S}{iv}$^{*}$ was allowed to adjust within a 3$\sigma$ error range during the fitting process.
%%%%%%%%%%%%%%%%%%%%%
We have checked the spectrum for any signs of the \ion{C}{iii}$^{*}$ $\lambda$ 1175\AA \ narrow absorption line and 
we were not able to identify 
any traces of the mentioned line. 
%To explain this, we  used Cloudy \citep{cloudy17} to predict the ionic column  density of two \ion{C}{iii} lines: first, \ion{C}{iii}  $\lambda$977\AA  \ which is not located in the wavelength range of the available dataset and second, \ion{C} {iii}$^{*}$ $\lambda$1175\AA \ which we could not identify in  our spectrum. Our results show that for  our absorption outflow system, $\frac{\textrm{N}_{ion}(\textrm{\ion{C}{iii}}^{*}\lambda 1175\AA) {\textrm{N}_{ion} (\textrm{\ion{C}{iii}}\  \lambda 977\AA)}$ is of order of 10$^{-4}$ (to perform these calculations, we used the electron number density, which is calculated in section~\ref{subsec:eden}). Given that the oscillator strength of \ion{C} {iii} is almost four times larger than that of \ion{C}{iii}$^{*}$, we expect $\frac{\tau(\ion{C}{iii}^{*})}{\tau(\ion{C}{iii})}$<0.001. This suggests that the presence of a very strong \ion{C}{iii} narrow absorption line would still indicate that \ion{C}{iii}$^{*}$ would be very shallow and not observable.  % The same Cloudy calculations estimate an optical depth of several thousand for \ion{C}{iii} $\lambda$ 977\AA, 
%and an optical depth
%<0.1 for \ion{C}{iii}$^{*}$ $\lambda$ 1175\AA . This means that this line would be very shallow and undetectable in the spectrum.
Table~\ref{tab1} summarizes the results we got from our N$_{ion}$ measurements, along with the numbers we chose to use in the next step, which will assess the outflow characteristics. The adopted errors include the corresponding AOD errors and a systematic error (20$\%$ of the adopted value), which are quadratically added together \citep[e.g.][]{xu18,mill18,mill20c}{}.

%%%%%%%%%%%%%%%%%%%%%

%%%%%%%%%%%%%%%%%%%%%%%%%%%%%
%%%%%%%%%%%%%%%%%%%%%%%%%%%
%\subsubsection{\ion{C}{iii}$^{*}$ $\lambda$ 1175\AA}

%%%%%%%%%%%%%%%%%%%%%%%%
\begin{center}
\begin{table}
\setlength{\tabcolsep}{8pt} % Default value: 6pt
\renewcommand{\arraystretch}{1.5}
\begin{tabular}{||c c c||} 
 \hline
 Ion & AOD & Adopted \\ [0.8ex] 
 \hline\hline
 \ion{C}{iv} & 580$^{+50}_{-40}$ & >580$_{-122}$  \\ 
 \hline
 \ion{Si}{iv} & 63$^{+7}_{-6}$ & >63$_{-14}$ \\
 \hline
 Ly$\alpha$ & 105$^{+2}_{-2}$ & >105$_{-21}$ \\
 \hline
 \ion{N}{v} & 503$^{+24}_{-31}$ & >503$_{-103}$ \\
 \hline
 \ion{C}{ii}  & 187$^{30}_{-27}$ & >187$_{-46}$ \\
 \hline
 \ion{Si}{ii} & 9$^{+1}_{-1}$ & 9$^{+1}_{-1}$ \\
 \hline
 \ion{Si}{iii} & 18$^{+1}_{-1}$ & >18$_{-4}$ \\
 \hline
 \ion{S}{iv} & 262$^{+31}_{-29}$ & 262$^{+61}_{-60}$ \\[1ex]
 \hline
 \ion{Si}{ii}$^{*}$ & 4$^{+1}_{-1}$ & 4$^{+1}_{-1}$ \\[1ex]
 \hline
\end{tabular}
\caption{The ionic column densities of the absorption lines detected in the J1429+4518 outflow system. Please note that the reported/adopted value for \ion{C}{ii} includes the column densities of \ion{C}{ii} and \ion{C}{ii}$^{*}$. We did not add the systematic error to \ion{Si}{ii} and \ion{Si}{ii}$^{*}$ since these two lines are only 5\AA\  apart and will not be affected much by a systematic error. All of the column density 
values are in units of 10$^{12}$ cm$^{-2}$.}\label{tab1}
\end{table}
\end{center}
%%%%%%%%%%%%%%%%%%%%%%%%%%%%%%%%%%%%%%
\subsection{Photoionisation Solution}
\label{sec:photosol}
\begin{figure}% [hpbt] what you need
    \centering
       %\subfigure{%{0.3\textwidth}
        \includegraphics[width=0.9\linewidth]{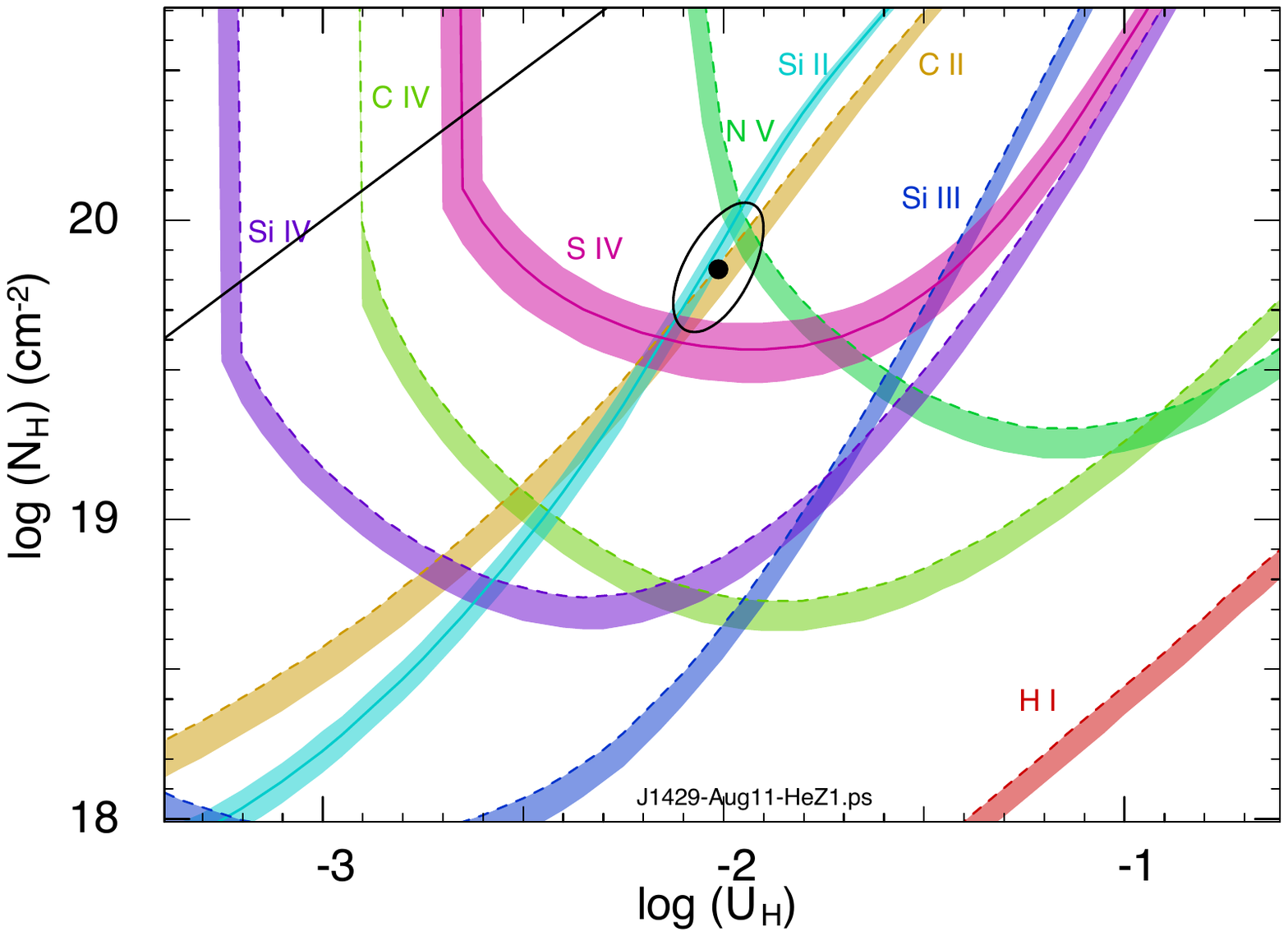}
        \label{fig5-a}
        % \subfigure{%
        \includegraphics[width=0.9\linewidth]{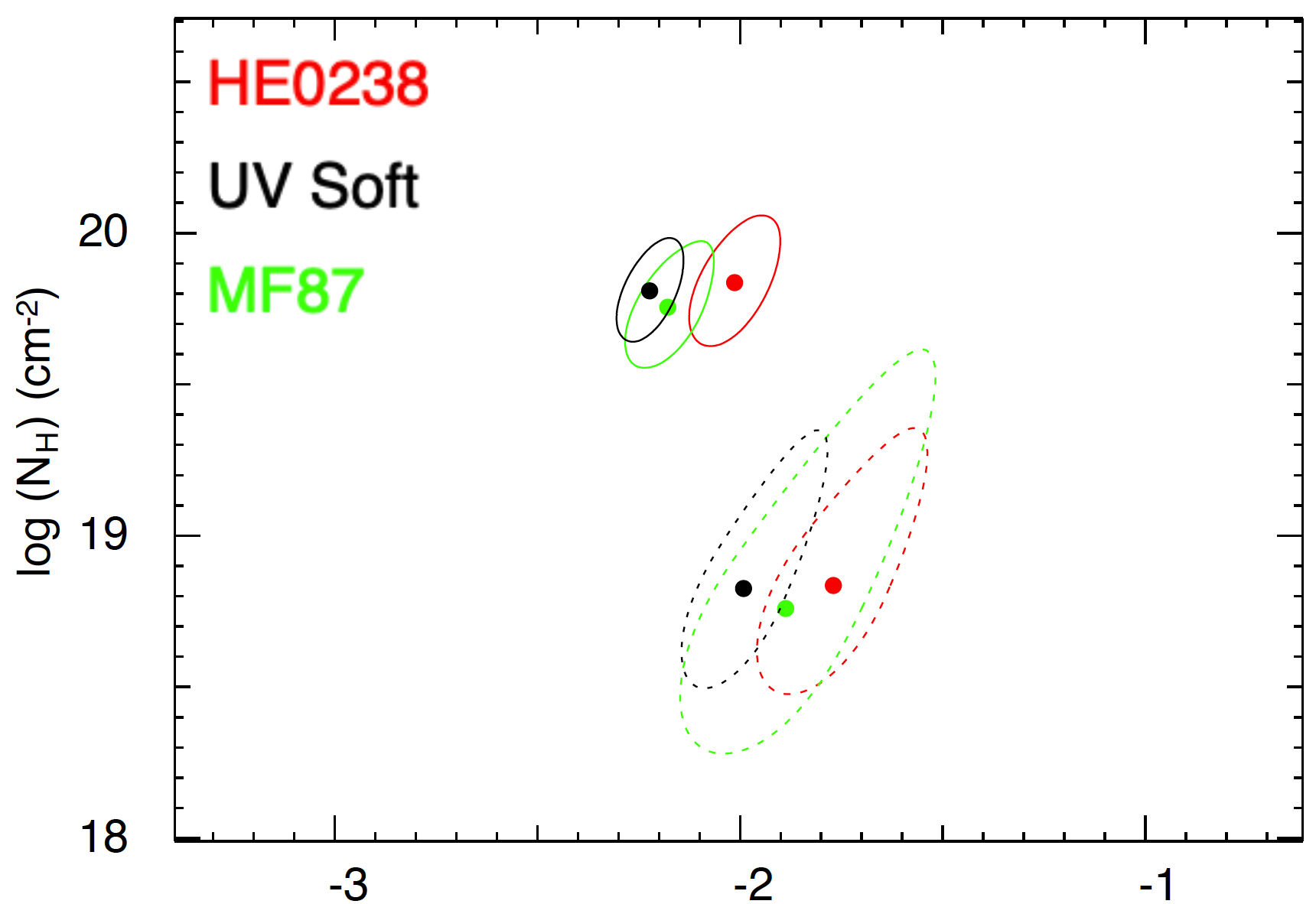}
        \label{fig5-b}
    \caption{Top panel: single phase photoionization solution for the absorption outflow system. Each coloured contour
    indicates the ionic column densities consistent with the observations (available in Table~\ref{tab1}), assuming the HE0238 SED and solar metallicity. Solid lines within the contours show the ionic column densities taken as measurements, while dashed lines contours belong to lower limits. The shaded bands are the uncertainties added for each contour. The black circle shows the best $\chi^{2}$-minimization solutions.
    Bottom Panel: The photoionization solution for a total of six models, including three SEDs (HE0238, MF87, and UV–soft) and two metallicities for each: solar metallicity and super-solar metallicity, for a total of six models.}\label{fig5}
\end{figure}

We calculated the ionic column densities of the mentioned absorption lines for the primary purpose of
estimating the characteristics of the outflow, 
including the total hydrogen column density (N$_{H}$) and the 
ionization parameter of the system (U$_{H}$) \citep[e.g.][]
{xu19,byun22a, byun22b, byun22c,walk22}. To do so, we 
use Cloudy simulations \citep{cloudy17} in which a 
grid of N$_{H}$ and U$_{H}$ are used to predict the 
abundance of various ions. Then we 
constrain the N$_{H}$ and U$_{H}$ values based on our 
measured ionic column densities. The top panel of 
Fig.~\ref{fig5} shows the results of this process.
To produce this Figure, we used the spectral energy 
distribution (SED) of AGN HE0238-1904 
\citep[hereafter HE0238,][]{arav13} in Cloudy and set a 
large range of grids on N$_{H}$ and U$_{H}$. In  
Fig.~\ref{fig5}-top panel, the coloured contours show the values labeled 
"adopted" in Table~\ref{tab1}. As this Figure shows, a single-
phase solution is sufficient to satisfy the 
constraints from the ionic column densities. Based on 
these results, we narrowed down the column density-ionization space to a pair of 
N$_{H}$ and U$_{H}$ for the absorption outflow system: 
U$_{H}=$--2.0$^{+0.1}_{-0.1}$ and 
N$_{H}$=--19.84$^{+0.20}_{-0.20}$ [cm$^{-2}$] (reduced 
$\chi^{2}$ = 2.5). Please note that to produce these 
results, a solar abundance was assumed.

While the above solution satisfies all the constraints, we examined the dependency of the results on the SED and metalicity. For this purpose, we repeated the same process of prediction N$_{H}$ and U$_{H}$ using another two SEDs, namely the MF87 SED \citep{mat87}, and UV-soft SED \citep{dun10}. For all three SEDs, we have also investigated the effects of super-solar metalicity 
(Z=4.68Z$_{\astrosun}$) \citep{ball08,mill20b}. The lower panel of Fig.~\ref{fig5} shows the effects of different SEDs and super-solar metalicity on the results. In this Figure, each solid line belongs to the calculations considering a solar metalicity while dashed lines are resulted from a super-solar metalicity. These results show that using MF87 and UV-soft SEDs results in a slightly smaller $\chi^{2}$ (2.3 and 1.5, respectively). However, as discussed by \cite{arav13}, we prefer to continue using HE0238 SED since it extends quite far into the extreme UV rest (EUV) wavelength range.
%%%%%%%%%%%%%%%%%%%%%%%%%%%%%%%%%%%%%%
%%%%%%%%%%%%%%%%%%%%%%%%%%%%%%%%%%%%%%
%%%%%%%%%%%%%%%%%%%%%%%
\subsection{Determining the electron number density and the distance of the outflow from the central source}
\subsubsection{Electron Number Density}
\label{subsec:eden}
\begin{figure}
\includegraphics[width=\columnwidth]{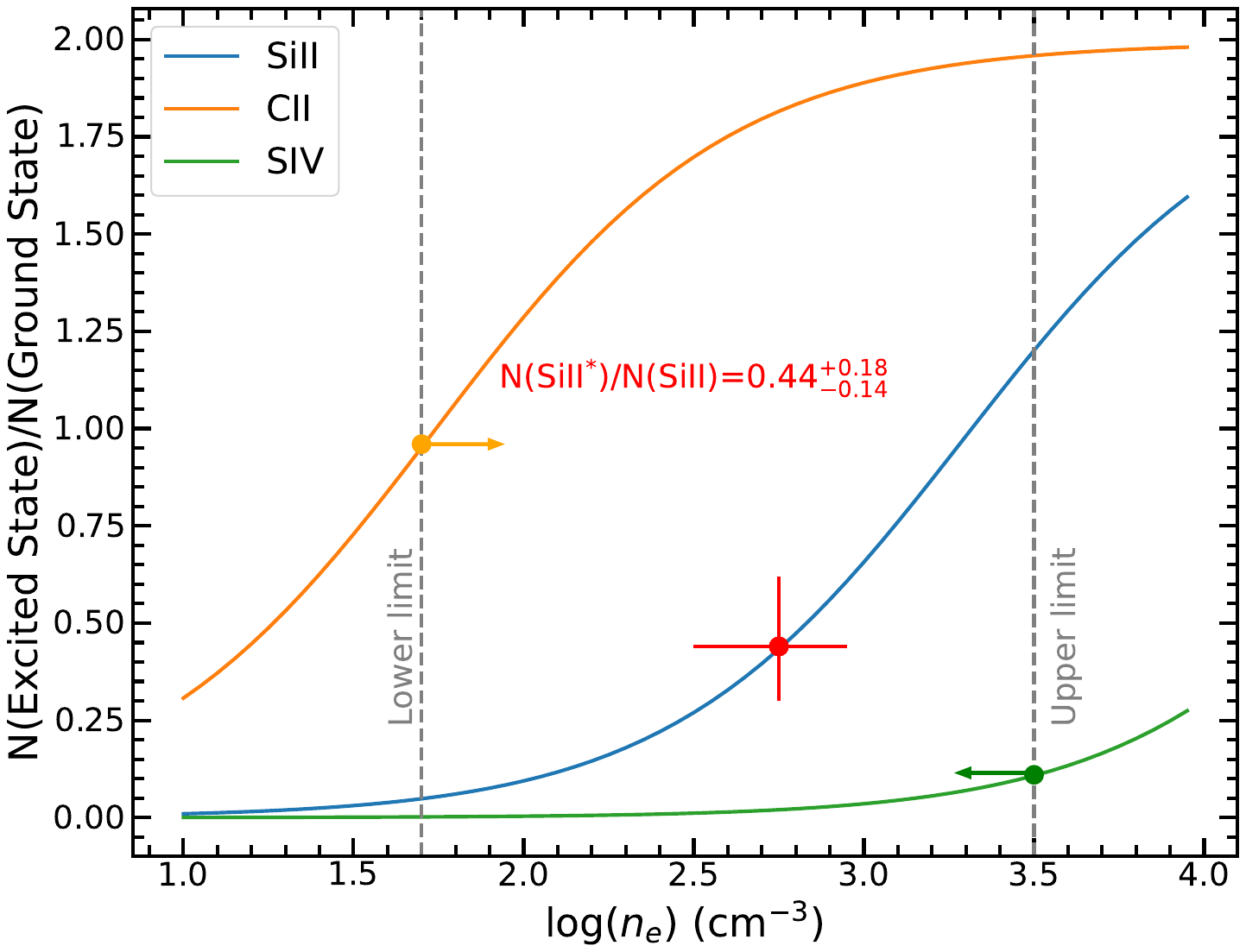}
      \caption{The excited state to resonance state column density ratio vs. the electron number density from  Chianti atomic database. We have shown the calculated value from our modeling and marked its corresponding electron number density by a vertical dashed line. Please note that these results are sensitive to the temperature, and to produce this plot, we assumed T=14000 K. }
         \label{fig6}
\end{figure}

We ran a model of Cloudy using HE0238 SED and the N$_{H}$ and U$_{H}$ values to predict the outflow's temperature. This simulation resulted in a temperature of T$\approx$14000 K. Assuming this temperature, we used Chianti 9.0.1 atomic database \citep{dere97, dere19} to estimate the abundance ratios of excited state to the resonance state for \ion{Si}{ii}, \ion{C}{ii}, and \ion{S}{iv} as a function of electron number density (n$_{e}$). Fig.~\ref{fig6} illustrates the results. Since \ion{Si}{ii}'s absorptions in both excited and resonance states were reliably measured (section~\ref{subsec:si2}), we took its ratio as a diagnostics for the electron number 
density of the outflow (blue curve in Fig. ~\ref{fig6}). These considerations result in $\log$~n$_{e}$= 2.75$^{+0.20}_{-0.25}$ [cm$^{-3}$]. As described in \cite{oster06} and for a highly ionized plasma, n$_{e}$$\approx$1.2n$_{H}$, resulting in $\log$~n$_{H}$= 2.67$^{+0.20}_{-0.25}$ [cm$^{-3}$]. We have also used \ion{C}{ii} and \ion{S}{iv} to predict lower and higher limits for n$_{e}$ (details in sections~\ref{subsec:c2} and ~\ref{subsec: s4}, respectively). These upper and lower constraints on n$_{e}$ are fully consistent with the \ion{Si}{ii}$^{*}$/\ion{Si}{ii} measurement.
%%%%%%%%%%%%%%%%%
%%%%%%%%%%%%%%%%%%
\subsubsection{Location of the Outflow}
Once N$_{H}$, n$_{H}$, and U$_{H}$ are known, we can use equation 14.4 from \cite{oster06} to extract the location of the outflow:
%%%%%%%%%%%%%%%%%%%%%%%%%
\begin{equation}
U_H\equiv\frac{Q(H)}{4\pi R^{2}c~n_{H}} \Rightarrow R=\sqrt{\frac{Q(H)}{4\pi c~n_{H} U_{H}}} \label{eq1}
\end{equation}
%%%%%%%%%%%%%%%%%%%%%%%
\noindent In which Q(H) [s$^{-1}$] is the number of hydrogen-ionizing photons emitted by the central object per second, R is the distance between the outflow
and the central source, n$_H$ is the hydrogen density and c
is the speed of light. 

To calculate R, we must first calculate the number of hydrogen-ionizing photons Q(H). To do so, we followed several steps explained below (please refer to the works done by \cite{mill20a}, \cite{byun22a,byun22b}, and \cite{walk22}):
%%%%%%%%%%%%%%%%%%%%%%
\begin{itemize}
\item First, We scaled the SED of HE0238 to match the continuum flux of J1429+4518 at observed wavelength of $\lambda$=1350\AA. Based on the HST observations we are using in this paper, at the mentioned wavelength F$_{\lambda} \approx 3 \times 10^{-15}$ erg s$^{-1}$ cm$^{-2}$ \AA$^{-1}$.
\item Then, we integrated over the scaled SED for all energies above 1 Ryd (Hydrogen ionization potential) to get a value of 1.2$^{+0.1}_{-0.1}\times 10^{54}$ s$^{-1}$ for Q(H) and a bolometric luminosity of L$_{bol}$= 2.13$^{+0.2}_{-0.2}\times 10^{44}$ erg s$^{-1}$. 
\item Finally, using these values and equation~\ref{eq1}, we predict that the outflow system is located at a distance of R$\approx $275$^{+53}_{-46}$ pc from the source.  
\end{itemize}
%%%%%%%%%%%%%%%%%%%%%%%%
\subsection{Black hole mass and outflow's energetics}
\label{sec:cal}
\subsubsection{Black Hole mass and the Eddington luminosity}
To get an estimate of the outflow's energetics, including its mass-loss rate and the kinetic luminosity, we need to get a hand on the mass of central BH and its Eddington luminosity first. 

\begin{figure}
\includegraphics[width=\columnwidth]{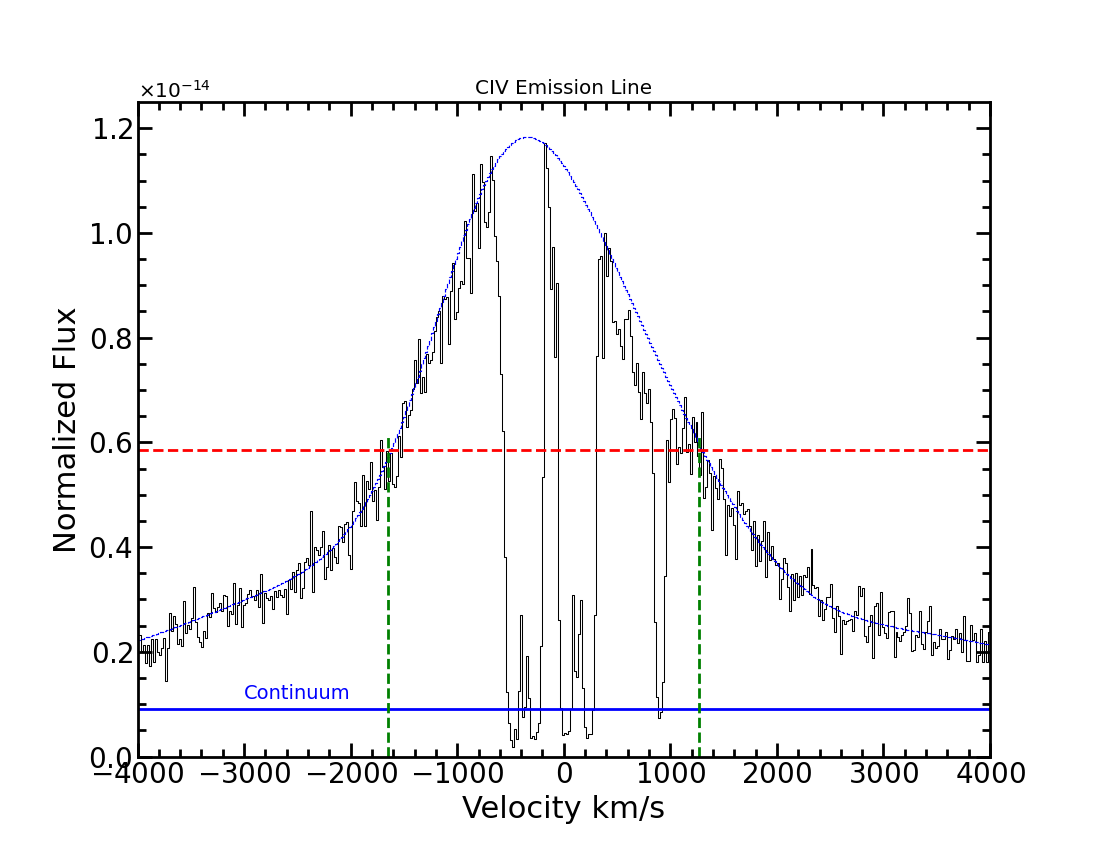}
      \caption{\ion{C}{iv} emission line region. The blue line shows our Gaussian fit to the emission spectra. The Red dashed line indicates where the half maximum is, while the green dashed lines show the full width of the half maximum. The continuum level is shown in the Figure, and it is subtracted for the purpose of FWHM determination.} 
         \label{fig7}
\end{figure}

In 2006, \citeauthor{vest06} showed that \ion{C}{iv} emission line could be used to measure the mass of the BH via this equation:
%%%%%%%%%%%%%%%%%%%%%%%%
\begin{equation}
\log M_{BH}(\textrm{\ion{C}{iv}})=\log ( {[\frac{\textrm{FWHM(\ion{C}{iv})}}{1000  \textrm{kms}^{-1}}]^{2}[\frac{\lambda L_{\lambda}(1350\AA)}{10^{44} \textrm{ergs}^{-1}}]^{0.53}}) \label{eq2}
\end{equation}
%%%%%%%%%%%%%%%%%%%%%%%%
Later in 2017, \citeauthor{coat17} revised equation~\ref{eq2} as below:
%%%%%%%%%%%%%%%%%%%%%%%%
\begin{equation}
    \begin{aligned}
M_{BH}(\textrm{\ion{C}{iv}, Corr.})=10^{6.71}[\frac{\textrm{FWHM(\ion{C}{iv}, Corr.)}}{1000 \textrm{kms}^{-1}}]^{2}\times \\
[\frac{\lambda L_{\lambda}(1350\AA)}{10^{44} \textrm{ergs}^{-1}}]^{0.53}
    \end{aligned}\label{eq3}
\end{equation}
%%%%%%%%%%%%%%%%%%%%%%%%%%%%%%%%
\noindent where:
%%%%%%%%%%%%%%%%%%%%%%%%%%%%%
\begin{equation}
    \begin{aligned}
\textrm{FWHM(\ion{C}{iv}, Corr.)}=\frac{\textrm{FWHM(\ion{C}{iv}, Meas.)}}{(0.41\pm0.02)\frac{\textrm{\ion{C}{iv} blueshift}}{1000 \textrm{kms}^{-1}}+(0.62\pm0.04)}
    \end{aligned} \label{eq4}
\end{equation}
%%%%%%%%%%%%%%%%%%%%%%%%%%%%%%%%
As shown in Fig.~\ref{fig2}, we can easily identify the \ion{C}{iv} emission and use its FWHM to achieve the mass of the black hole. Fig.~\ref{fig7} shows our Gaussian model to fit the spectra at \ion{C}{iv} 
region, in which we have modeled the \ion{C}{iv} emission with a single Gaussian. Based on this Figure, the FWHM is measured to be $\approx$3100 km~s$^{-1}$ for the \ion{C}{iv} emission line. We also measured an average flux of 
1$\times$10$^{-15}$ erg s$^{-1}$ cm$^{-2}$ \AA$^{-1}$ for $\lambda_{\textrm{rest}}$=1350\AA, which for our narrow outflow system results in $\lambda$L$_\lambda$=1.16$\times$10$^{42}$ erg s$^{-1}$. Using these values in the equation~\ref{eq4} $\&$ \ref{eq3}, we get a black hole mass of M$_{BH}$=7.5$\times$10$^{6}$M$_{\astrosun}$, resulting in an Eddington luminosity of L$_{Edd}$=8.8$\times$10$^{44}$ erg~s$^{-1}$.
%%%%%%%%%%%%%%%%%%%%%%%%
%%%%%%%%%%%%%%%%%%%%%%%%
\subsubsection{Outflow's Energetics}
Using outflow's R and N$_{H}$ obtained above, we can estimate the mass-loss rate and the kinetic luminosity of the outflow using the following equations\citep{borg12b}:
%%%%%%%%%%%%%%%%%%%%%%%%%%%%%%%%%%
\begin{equation}
\Dot{M} \simeq 4\pi \Omega R N_H \mu m_{p} \nu \label{eq5}
\end{equation}
%%%%%%%%%%%%%%%%%%%%%%%%%%%%%%%%%%
\begin{equation}
\Dot{E}_{K} \simeq \frac{1}{2}\Dot{M} v^{2}\label{eq6}
\end{equation}
%%%%%%%%%%%%%%%%%%%%%%%%
\noindent In equation~\ref{eq5}, $\mu$ = 1.4 is the mean atomic
mass per proton, v is the velocity of the outflow, and m$_{p}$ is the mass of a proton \citep{borg12a}. 
In the mentioned equation, $\Omega$ is the global covering factor defined as the percentage of the source covered by the outflow in all directions (not only line-of-sight). We assume $\Omega=0.5$ as \cite{misa07} explain that intrinsic NALs are observed in at least 50$\%$ of AGNs. In an earlier study published in 2003, \citeauthor{vest03} discusses more than half of the AGNs embed NALs. All these studies are in agreement with the 50-70$\%$ population ratio of NALs among Seyfert 1 galaxies as reported by \cite{cren99}.  
 Assuming all these values and solving equations~\ref{eq5} and \ref{eq6} for the mass-loss rate and the kinetic luminosity results in $\Dot{M}$=0.22$^{+0.09}_{-0.06}$$M_{\astrosun} yr^{-1}$ and $\log \Dot{E_{K}}$=39.3$^{+0.1}_{-0.2}$ [erg s$^{-1}$]. Table~\ref{tab2} summarizes all these results along with the results from section~\ref{sec:anal}.

As reported in Table~\ref{tab2}, the outflow system's kinetic luminosity ratio to the source's Eddington luminosity is $\approx 0.0002 \%$. \cite{hop10} explain that an outflow system must have a $\Dot{E_{K}}/{L_{Edd}}$ of at least $\approx 0.5 \%$ to contribute to the AGN outflow, verifying that the NAL outflow system identified here does not contribute to AGN feedback processes.

It is worth mentioning that if we take the redshift considerations into account  (Section \ref{subsec:red}) and assume that the outflow has a velocity of approximately --195 km~s$^{-1}$, then $\Dot{M}$=0.27$M_{\astrosun} yr^{-1}$ and $\log \Dot{E_{K}}$=39.5[erg s$^{-1}$]. Based on these values,  the outflow system's kinetic luminosity ratio to the source's Eddington luminosity would be $\approx 0.0004 \%$. This indicates that for the [OIII]-based redshift, the above discussion remains valid, and this outflow contributes minimally or not at all to the AGN feedback process.
%%%%%%%%%%%%%%%%%%%%%%%%
\begin{center}
\begin{table}
\setlength{\tabcolsep}{10pt} % Default value: 6pt
\renewcommand{\arraystretch}{1.5}
\begin{tabular}{||c c||} 
 \hline
 %Physical Properties of the   NAL Outflow system  & \\ [0.8ex] 
 \hline
 $\log N_{H}$ [cm$^{-2}$] & 19.84$^{+0.20}_{-0.20}$  \\ 
 \hline
 $\log U_{H}$ [dex] & -2.0$^{+0.1}_{-0.1}$  \\
 \hline
 $\log n_{e}$ [cm$^{-3}$] & 2.75$^{+0.20}_{-0.20}$ \\
 \hline
 R [pc] & 275$^{+53}_{-46}$ \\
 \hline
 $\Dot{M}$ [$M_{\astrosun} yr^{-1}$] & 0.22$^{+0.09}_{-0.06}$ \\
 \hline
 $\log \Dot{E_{K}}$ [erg s$^{-1}$] &  39.3$^{+0.1}_{-0.2}$\\
 \hline
 $\Dot{E_{K}}/{L_{Edd}}$ & 2.0$^{+0.9}_{-0.7}$ $\times$10$^{-6}$ \\
 \hline
 $\Dot{E_{K}}/{L_{Bol}}$ & 8.5$^{+3.0}_{-2.0} $$\times$10$^{-6}$ \\[1ex] 
 \hline
\end{tabular}
\caption{Physical properties of the narrow absorption line outflow system }\label{tab2}
\end{table}
\end{center}
%%%%%%%%%%%%%%%%%%%%%%%%
%%%%%%%%%%%%%%%%%%%%%%%%
\section{Discussion}
\label{sec:disc}
\subsection{The thermal stability curve}
When gas is photoionized, it will reach the equilibrium defined by an ionization parameter \citep{tart69, kall01}:

\begin{equation}
\xi\equiv\frac{L_{\rm ion}}{n_{H} r^{2}} \label{eq7} [\textrm{erg~cm~s}^{-1}]  
\end{equation}
\noindent and a specific temperature. This equilibrium arises due to the interplay between various heating and cooling mechanisms that depend on the gas's physical and chemical attributes and the characteristics of the intrinsic radiation of the source.

\begin{figure}
\includegraphics[width=\columnwidth]{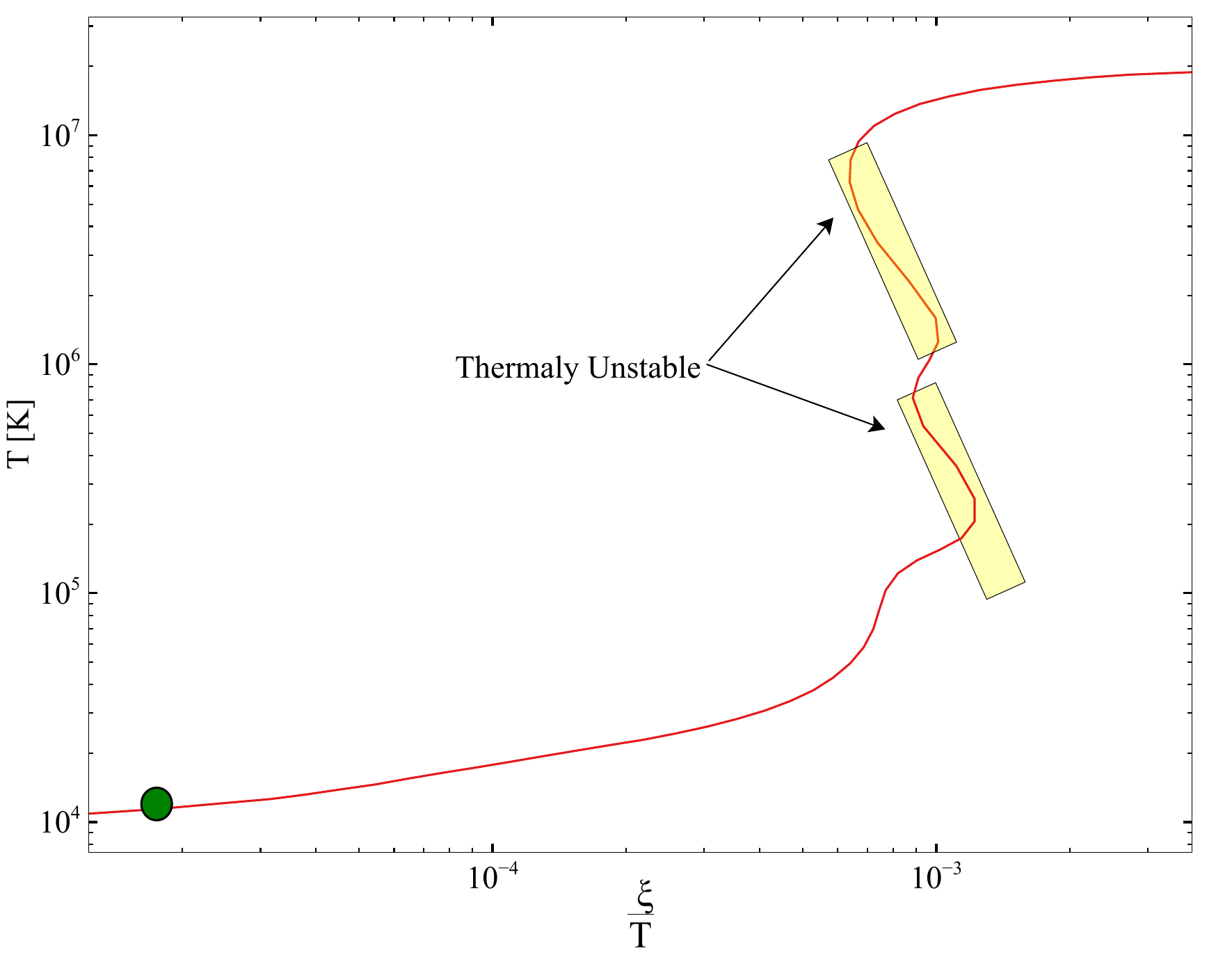}
      \caption{The photoionization stability curve (s-curve) for J1429+4518. The green circle indicates the narrow absorption outflow system discussed in this paper. Yellow rectangles indicate the negative-slope regions, meaning any gas with those characteristics would be unstable. }
         \label{fig8}
\end{figure}

It is common to represent these balanced states on a stability curve, depicted in a diagram with temperature (T) on the vertical axis and the ratio of $\xi$/T on the horizontal axis (for 
example, \cite{krol81}). When we assume that L/n(H)r$^{2}$ remains constant , the latter parameter, $\frac{\xi}{T}$ $\propto$ $\frac{p_{rad}}{p_{gas}}$, indicates the gas pressure. 
The thermal stability curve (S-curve) for J1429+4518 is produced using Cloudy and is shown in Fig.~\ref{fig8}. To create the stability curve, we have used the previously mentioned HE0238 SED and the outflow's characteristics are adopted from Table~\ref{tab2}. In this Figure, the region on the left-hand side of the curve is dominated by cooling processes, while heating processes dominate the right-hand side. A positive slope on the S-curve happens when an increase in T (due to an isobaric perturbation)  will move the gas to the cooling region and so revert the gas to the same equilibrium state. Such equilibrium states are considered to be thermally stable. The green circle in Fig.~\ref{fig8} indicates where our narrow absorption line outflow is located regarding the stability. Based on this plot, the outflow is on the positive slope and hence it is assumed to be thermally stable. 
%%%%%%%%%%%%%%%%%%%%%%%%
%%%%%%%%%%%%%%%%%%%%%%%%
\subsection{The Importance of the NALs}
\cite{ham11} explain that to understand AGN outflows better, we need clearer information about the different types of outflows, including NAL outflows. For a while now, there has been an ongoing discussion about whether the NALs and BALs (Broad Absorption Lines) have any connection or relationship \citep{wey79}. Furthermore, there's an ongoing debate about how these associated NALs and BALs might fit into the overall development or evolution of AGNs \citep{brig84,ham01}.

It's not straightforward to detect NAL outflows because they're hard to identify: most narrow lines we see in AGN spectra come from unrelated cosmic gas, not from the AGN themselves. However, some studies suggest that a significant part of these narrow absorption lines actually come from AGN outflows \citep[e.g.][]{nest08, wild08, rich01}. 
%%%%%%%%%%%%%%%%%%%
\subsection{Similarity to Milky Way Outflows}
The narrow absorption line outflow system in J1429+4518 shows some remarkable similarity to the two bipolar lobe outflows observed in the Milky Way by XMM-Newton
and Chandra (see \cite{pont19}, figure 1). One of these lobe outflows extends up to 160 pc from the Galactic Center (GC) in the northern direction. The second lobe, situated to the south, extends even further, reaching beyond the southern Fermi Bubble's base and spanning approximately 250 pc \citep{veil20}. Notably, the distances from their central source align closely with the 275 pc measured for our study's
NAL outflow system. Furthermore, their kinematic luminosities are estimated at $\log \Dot{E_{K}}$=39.6 [erg s$^{-1}$] \citep{pont19}---merely 0.3 dex larger than that of our outflow system—these similarities are striking. Finally, the comparability of the Milky Way's black hole mass (M$_{BH}$=4.15$\times$10$^{6}$M$_{\astrosun}$; \cite{grav19}) with that of J1429+4518 (M$_{BH}$=6.7$\times$10$^{6}$M$_{\astrosun}$) along with the mentioned similarities in the outflow's measurements, suggest a common origin for the two outflows. It's worth noting that other outflows in the Milky Way are located farther away or have much larger kinetic luminosity\citep{veil20}.
%%%%%%%%%%%%%%%%%%%%%%%%%%%%%%%%%
\section{Conclusions}
\label{sec: con}
In this paper, we analysed two datasets from HST/COS 2021 observations of 
 Seyfert 1 galaxy J1429+4518. We successfully identified a narrow absorption line outflow system located $\sim$275 pc away from the central source. This outflow was identified through narrow absorption troughs of \ion{C}{iv}, \ion{Si}{iv}, \ion{N}{iv}, Ly$\alpha$, \ion{C}{ii}, \ion{C}{ii}$^{*}$, \ion{Si}{ii}, \ion{Si}{ii}$^{*}$, \ion{Si}{iii}, and \ion{S}{iv}. We measured the centroid velocity of the outflow system to be --151 km~s$^{-1}$ and extracted the ionic column densities of the absorption lines using the AOD method. 

As the next step, measured ionic column densities were used to determine the total hydrogen column density ($\log$ N$_H$=19.84 erg~s$^{-1}$) and the ionization state ($\log$ U$_H$=-2.0) of the NAL outflow. This was done by using $\chi^{2}$-minimization solutions and performing Cloudy simulations. We also used the ratio of the ionic column densities of \ion{Si}{ii}$^{*}$ to \ion{Si}{ii} for measuring the electron number density ($\log$ n$_e$=2.75 cm$^{-3}$) and hence its hydrogen density. Having the hydrogen density enabled us to estimate the outflow location mentioned above. We followed all these calculations by estimating the energetics of the outflow: this outflow has a kinetic luminosity of $\log \Dot{E_{K}}$=39.3[erg s$^{-1}$] and its mass-loss rate is measured to be $\Dot{M}$=0.22$M_{\astrosun} yr^{-1}$. 

The NAL outflow of J1429+4518 is thermally stable, but it does not contribute to AGN feedback processes since its kinetic luminosity is only $\approx 0.00025 \%$ of the Eddington luminosity of the central source. Finally, we note that this NAL outflow exhibits a striking similarity to the two bipolar lobe outflows detected in the Milky Way through observations conducted by XMM-Newton and Chandra. The similarities in their energetics and location, and having central black holes with similar masses, strongly suggest that these outflows might have a common origin.

%%%%%%%%%%%%%%%%%%%%%%%%
%%%%%%%%%%%%%%%%%%%%%%%%
%%%%%%%%%%%%%%%%%%%%%%%%
%%%%%%%%%%%%%%%%%%%%%%%%
%%%%%%%%%%%%%%%%%%%%%%%%
%%%%%%%%%%%%%%%%%%%%%%%%
%%%%%%%%%%%%%%%%%%%%%%%%
%%%%%%%%%%%%%%%%%%%%%%%%

\section*{Acknowledgements}
We express our appreciation to the anonymous reviewer whose feedback has contributed to the enhancement of this manuscript. We acknowledge support
from NSF grant AST 2106249, as well as NASA STScI grants AR-
15786, AR-16600, AR-16601, and HST-AR-17556. 

%%%%%%%%%%%%%%%%%%%%%%%%%%%%%%%%%%%%%%%%%%%%%%%%%%
\section*{Data Availability}
The data of J1429+4518 described in this paper may be obtained
from the MAST archive at \url{https://dx.doi.org/10.17909/mzeb-fm44}


\bibliographystyle{mnras}

%\begin{thebibliography}{99}
%\end{thebibliography}

%%%%%%%%%%%%%%%%%%%%%%%%%%%%%%%%%%%%%%%%%%%%%%%%%%

%%%%%%%%%%%%%%%%% APPENDICES %%%%%%%%%%%%%%%%%%%%%

\appendix

%%%%%%%%%%%%%%%%%%%%%%%%%%%%%%%%%%%%%%%%%%%%%%%%%%

% Don't change these lines
\bsp	% typesetting comment
\label{lastpage}
\end{document}